\documentstyle[12pt,epsfig]{article}
\newlength{\dinwidth}
\newlength{\dinmargin}
\newcommand\ben{\begin{equation}}
\newcommand\een{\end{equation}}
\newcommand\bea{\begin{eqnarray}}
\newcommand\eea{\end{eqnarray}}

\newcommand\vx{{\vec x}}
\newcommand\vy{{\vec y}}

\newcommand\nn{\nonumber}

\newcommand\tg{\tilde{g}}
\newcommand\tR{\tilde{R}}
\newcommand\bT{{\bar{T}}}

\begin{document}
\thispagestyle{empty}
\addtocounter{page}{-1}
\vskip-0.35cm
\begin{flushright}
UK/07-11 \\
\end{flushright}
\vspace*{0.2cm} \centerline{\Large \bf Gauge Theory Duals of Cosmological}
\centerline{\Large \bf Backgrounds and their Energy
Momentum Tensors}
\vspace*{1.0cm}
\centerline{\bf Adel Awad${}^{a,b}$\footnote{On leave of absence from  
Ain Shams University, Cairo, EGYPT}, Sumit R. Das${}^a$, 
K. Narayan ${}^c$ and Sandip P. Trivedi ${}^d$}
\vspace*{0.2cm}
\centerline{\it Department of Physics and Astronomy,}
\vspace*{0.2cm}
\centerline{\it University of Kentucky, Lexington, KY 40506 \rm USA ${}^a$}
\vspace*{0.35cm}
\centerline{\it Center for Theoretical Physics, British University of
  Egypt}
\vspace*{0.2cm}
\centerline{\it Sherouk City 11837, P.O. Box 43, EGYPT ${}^b$}
\vspace*{0.35cm}
\centerline{\it Chennai Mathematical Institute,}
\vspace*{0.2cm}
\centerline{\it Padur PO, Siruseri 603103, \rm INDIA ${}^c$}
\vspace*{0.35cm}
\centerline{\it Tata Institute of Fundamental Research,}
\vspace*{0.2cm}
\centerline{\it Mumbai 400005, \rm INDIA ${}^d$}
\vspace*{1.0cm}
\centerline{\tt adel@pa.uky.edu, das@pa.uky.edu}
\centerline{\tt
narayan@cmi.ac.in, sandip@theory.tifr.res.in}

\vspace*{0.8cm}
\centerline{\bf Abstract}
\vspace*{0.3cm}
\vspace*{0.5cm}
\baselineskip=18pt
We revisit Type IIB supergravity backgrounds with null and spacelike
singularities with natural gauge theory duals proposed in {\tt
hep-th/0602107} and {\tt hep-th/0610053}. We show that for these
backgrounds there are always choices of the boundaries of these deformed
$AdS_5 \times S^5$ space-times, such that the dual gauge theories live
on {\it flat} metrics and have space-time dependent couplings. We
present a new time dependent solution of this kind where the effective
string coupling
is always bounded and vanishes at a spacelike singularity in the bulk,
and the
space-time becomes $AdS_5 \times S^5$ at early and late times.
The holographic energy momentum tensor calculated with a
choice of flat boundary is shown to vanish for null backgrounds and
to be generically non-zero for time dependent backgrounds.

\newpage

\section{Introduction}

In two previous papers \cite{Das:2006dz,Das:2006pw} three of us
proposed gauge theory duals to a class of time dependent and null
backgrounds of IIB supergravity. These solutions are deformations of
$AdS_5 \times S^5$ backgrounds with non-normalizable modes of the
metric and the dilaton. The null solutions and their duals were also
proposed in \cite{Chu:2006pa}. It is thus natural to conjecture that
the dual gauge theory is deformed by corresponding sources. Generally,
the supergravity solutions are singular with spacelike or null
singularities where of course supergravity breaks down. The idea is to
investigate whether the dual gauge theory remains well behaved in this
region and possibly provides a way to continue the time evolution
beyond the time where the supergravity is singular. Other discussions
of similar solutions include 
\cite{Sfetsos:2005bi}. For other approaches to the use of AdS/CFT
correspondence to study time dependent backgrounds, see 
\cite{Hertog:2004rz}.

These solutions have an Einstein frame metric of the form (with the
AdS scale $R_{AdS}=1$)
\ben ds^2 = \frac{1}{z^2} \left[ dz^2 +
\tg_{\mu\nu}(x) dx^\mu dx^\nu \right] + d\Omega_5^2
\label{one}
\een
and a dilaton $\Phi(x)$ together with a 5-form field strength
\ben
F_5 = \omega_5 + \star \omega_5
\een
This is a solution if
\ben
\tR_{\mu\nu}  =  \frac{1}{2}\partial_\mu \Phi \partial_\nu \Phi ,
~~~~~~~~~~~~\nabla^2 \Phi =  0
\label{three}
\een
where $\tR_{\mu\nu}$ is the Ricci tensor for the four dimensional
metric $\tg_{\mu\nu}(x)$.

In this paper, the $S^5$ part of the metric will remain unaltered and
we will not write this out explicitly.

In these coordinates the boundary is at $z=0$, and as argued in
\cite{Das:2006dz,Das:2006pw}, it is reasonable to assume that the dual
gauge theory lives on a 3+1 dimensional spacetime with metric
$\tg_{\mu\nu}(x)$ and has a spacetime dependent coupling $g_{YM} (x)  =
e^{\frac{\Phi(x)}{2}}$. Of particular interest are solutions where
\ben \tg_{\mu\nu}(x)dx^\mu dx^\nu = e^{f(x)}\left[ -2 dx^+ dx^- +
d\vx^2 \right]
\label{two}
\een
Now, because of the spacetime dependence of the coupling constant,
the Yang-Mills theory is not conformally invariant in the sense of
conformal coordinate transformations. However, the theory is still Weyl
invariant under Weyl transformations of the metric and corresponding
transformations of the fields. One might therefore hope that the
overall factor in (\ref{two}) can be removed by a Weyl transformation
leaving us with a gauge theory on flat space with spacetime dependent
coupling. Such a step would be, however, subtle in the quantum theory
because of a possible Weyl anomaly.

Null solutions with the conformal factor depending on a single null
coordinate are of special interest from several points of view.
First,
for such solutions, with $f = f(x^+)$ and $\Phi = \Phi (x^+)$ the
equation (\ref{three}) becomes
\ben
\frac{1}{2}(f^\prime)^2 - f^{\prime\prime} = \frac{1}{2}(\Phi^\prime)^2
\label{four}
\een
This means that we have a one function-worth of solutions. We can pick
any $f(x^+)$ and solve for $\Phi (x^+)$. In particular, we may look
for solutions where the dilaton is bounded everywhere. Indeed an
interesting solution is given by
\ben
e^{f(x^+)} = (\tanh~x^+)^2 ~~~~~~~e^{\Phi (x^+)} = g_s | \tanh (\frac{x^+}{2}) |^{\sqrt{8}}
\label{five}
\een
This solution asymptotes to $AdS_5 \times S^5$ with string coupling
$g_s$ at $x^+ \rightarrow \pm \infty$. Both the metric and the
effective string coupling $e^\Phi$ drops to zero at $x^+ =0$ which is
the location of the null singularity. This is good : things appears to
be controlled. What makes this kind of background appealing is
the fact that the Weyl anomaly in the gauge theory living in the
metric (\ref{two}) identically vanishes for such null
backgrounds. Therefore, we could perform a Weyl transformation without
bothering about the anomaly, and obtain a dual theory which is on flat
space. In this dual theory, the coupling is always bounded, vanishing
at $x^+ = 0$.  One would expect that the Yang-Mills theory is well
behaved for such profiles. A careful argument along the above lines
was given in \cite{Das:2006pw}.

This provides a clean formulation of the problem in terms of light
front evolution. Consider the strong 't Hooft coupling regime of the
gauge theory
\ben g_{YM}^2 = g_s \rightarrow 0~~~~~~~~N \rightarrow
\infty~~~~~~~~ g_{YM}^2 N = {\rm finite~~and~~large}
\een
Then at $x^+
\rightarrow -\infty$ we start with a gauge theory in its ground
state. Since the 't Hooft coupling is large, supergravity in $AdS_5
\times S^5$ provides an accurate description in this regime.  Now turn
on a source leading to $x^+$ dependence
of the effective coupling, $g_{YM}^2
\rightarrow g_{YM}^2~e^{\Phi(x^+)}$. In the dual supergravity this
means that we have turned on a {\em non-normalizable} dilaton mode
$\Phi (x^+)$. As $x^+$ increases, the system evolves. In the gauge
theory side the effective coupling decreases, but remains large so
long as $|x^+|$ is large enough. In this regime, therefore, the
supergravity description remains good and that is what is described by
the supergravity solution. As we approach $x^+ = 0$, the gauge theory
coupling becomes weak, becoming exactly zero at $x^+ = 0$. A gauge
theory with weak coupling is not well described by dual supergravity -
stringy effects are important. Indeed, if we look at the evolution on
the supergravity side, and extrapolate the solution to small $x^+$ -
where supergravity should no longer be valid - we encounter a
singularity. However, because the gauge coupling is weak, one would
imagine that the gauge theory description remains good - maybe even
well approximated by perturbation theory. This gauge theory then
describes the region which appears to be singular if the supergravity
solution is extrapolated to the regime where it shouldnt have been in
the first place. Because of null dependence of the coupling, the gauge
theory has several properties which indicate that the theory is
actually well behaved at $x^+=0$. A null isometry ensures absence of
particle production, and in a light cone gauge the kinetic terms are
standard so that only positive powers of the $x^+$ dependent coupling
appear in the nonlinear terms \cite{Das:2006pw}. Since the coupling in
fact vanishes at $x^+ = 0$ one might hope that perturbation theory is
reliable in this region and may be even used to extend the time
evolution through $x^+=0$ to positive $x^+$.

In contrast to the null solutions, the time dependent solutions found
in \cite{Das:2006dz} do not have such nice features. These solutions
have boundary metrics which are Kasner cosmologies, e.g.
\bea
ds^2 & = &
\frac{1}{z^2} \left[ dz^2 - dt^2 + \sum_{i=1}^3 t^{2p_i}dx^i dx^i
\right]~~~~~~~\sum_i p_i = 1 \nn \\ e^{\Phi(t)} & = &
|t|^{\sqrt{2(1-\sum p_i^2)}}
\label{ten}
\eea
The string coupling - and therefore the Yang-Mills coupling - still
goes to zero at the spacelike singularity at $t = 0$, but diverges at
early or late times. As we shall see below it still turns out that the
Weyl anomaly of the boundary theory in these coordinates
vanishes. However, because of the divergence of the Yang Mills
coupling, it is unclear whether the gauge theory makes sense.
Furthermore, as discussed below, time dependent backgrounds (unlike
null backgrounds) have curvature singularities at $z=\infty$ for any
time. For the solution (\ref{ten}), at any fixed time $t$ the Ricci
scalar diverges as $\frac{z^2}{t^3}$. At early and late times the
divergence, however, goes away.

While the results stated above are suggestive, there are several
causes for serious concern. The first issue concerns the choice of
boundary described above. In this choice, the null or space-like
singularity extends all the way to the boundary. While a Weyl
transformation removes this and brings us to a flat boundary metric,
the conformal factor required is clearly singular : this is certainly
an uncomfortable situation. Furthermore, both for the null and the
spacelike cases, the behavior of the Yang-Mills coupling is {\em
non-analytic} \cite{Shenker}, casting serious doubts about a smooth
time evolution through this point. In fact, it turns out that if the
conformal factor $e^{f(x^+)}$ is chosen to vanish at $x^+=0$ in an
analytic fashion, $e^{\Phi (x^+)}$ has this non-analytic behavior.  On
the other hand if $e^{\Phi (x^+)}$ is analytic, $e^{f(x^+)}$ becomes
non-analytic.  Finally, the Kasner like solutions with space-like
singularities do
not appear to lead to a controlled dual theory.

In this paper we take some steps in solving some of these problems. We
show that for the solutions with brane metrics conformal to flat
space, one can always choose a foliation such that the boundary is
flat. To show this, we use the well known fact that in asymptotically
AdS space-times, a Weyl transformation of the boundary theory
corresponds to a special class of coordinate transformations in the
bulk - the Penrose-Brown-Henneaux transformations. We find these
exact transformations for any null solution of this kind and for the
Kasner solutions described above. For other time dependent solutions,
we find these transformations in a systematic expansion around the
(new) boundary. Thus in these new coordinates the boundary theory is
explicitly defined on flat space and the nontrivial feature is in the
time (or null time) dependence of the coupling. Furthermore, for the
null solutions of the type (\ref{two}), we do not have to worry about
the function $f(x^+)$ any more and choose the function $\Phi (x^+)$ to
have a nice analytic behavior at $x^+ = 0$.

We then describe new supergravity solutions
with spacelike singularities which asymptote to $AdS_5 \times
S^5$ at early and late times. The couplings are bounded
everywhere and vanish at the spacelike singularity. This opens up the
possibility of posing questions about space-like singularities in the
gauge theory dual in a fashion analogous to our formulation of null
singularities. At this time, however, we are still unable to arrive at
such a clean formulation for spacelike singularities.

We then compute the energy momentum tensors of these solutions using
standard techniques of holographic
RG\cite{skenderis,balasubramanian,nojiri:2000,Emparan:1999pm,Awad:1999xx}. We
find that the energy momentum tensor vanishes for all null backgrounds
for both foliations. For time dependent backgrounds, the trace anomaly
vanishes in the coordinates of (\ref{one}), while the energy momentum
tensor in the foliation leading to a flat boundary metric is non-zero.
In fact the answer diverges at
the time when the coupling constant vanishes. While a nonzero energy
momentum may be interpreted as particle production, one cannot attach
any significance to the divergence since this happens at the place
where the supergravity description is invalid.

While this paper was being written, \cite{Chu:2007um} appeared on the
archive, which has some overlap with our Section 2.

\section{Null cosmologies in better coordinates}

In this section we will rewrite the supergravity solutions of
\cite{Das:2006dz} in new coordinates leading to a choice of the
boundary with a flat metric.

\subsection{PBH transformations}

It is well known that Weyl transformations in the boundary theory
correspond to special coordinate transformations in the bulk - the
Penrose-Brown-Henneaux (PBH) transformations
\cite{Penrose:1986ca,Brown:1986nw}. Any asymptotically AdS
space-time may be written in a standard coordinate system of the
Feffermann-Graham form \ben ds^2 =
\frac{1}{{\bar{\rho}}^2}d{\bar{\rho}}^2 +
\frac{1}{{\bar{\rho}}^2}{\tilde{g}}_{\mu\nu}(x,{\bar{\rho}})dx^\mu dx^\nu
\label{fgform}
\een Now consider
the coordinate transformations
\cite{Imbimbo:1999bj,skenderis,Fukuma:2002sb} \ben {\bar{\rho}}
\rightarrow {\bar{\rho}}~e^{-\sigma(x,{\bar{\rho}})} ~~~~~ x^\mu
\rightarrow x^\mu + a^\mu (x,{\bar{\rho}})
\een
which keeps this form of
metric invariant. For infinitesimal transformations this is
ensured by requiring $\sigma$ to be a function of $x$ alone, and
\ben
\frac{1}{{\bar{\rho}}}\partial_{\bar{\rho}} a^\mu = - {\tilde{g}}^{\mu\nu}
\partial_\nu \sigma
\een 
The transformation of the metric ${\tilde{g}}_{\mu\nu}$ is given by
\ben
\delta {\tilde{g}}_{\mu\nu} (x,{\bar{\rho}}) = 2\sigma (x,{\bar{\rho}})
(1-\frac{1}{2} {\bar{\rho}}
\partial_{\bar{\rho}}) {\tilde{g}}_{\mu\nu} (x,{\bar{\rho}}) + 
\nabla_{(\mu} a_{\nu)} (x,{\bar{\rho}})
\label{six} \een The expression (\ref{six}) explicitly shows that
this transformation includes a Weyl transformation of the metric
${\tilde{g}}_{\mu\nu}$.

Consider now a metric of the form
\ben
ds^2 = \frac{1}{z^2} \left[ dz^2 + e^{f(x)}\eta_{\mu\nu}dx^\mu dx^\nu \right]
\een
Our aim is to perform a PBH transformation to remove the conformal
factor $e^{f(x)}$ in the boundary metric. However we need to do this
for finite PBH transformations.

\subsection{Null cosmologies in new coordinates}

When the conformal factor $f(x)$ is a function of a single null
coordinate $x^+$, i.e. when the original metric is of the form
\ben
ds^2 = \frac{1}{z^2} \left[ dz^2 + e^{f(x^+)}(-2dx^+ dx^- + d\vx^2
)
\right]
\een
it turns out that it is easy to figure out the correct
finite PBH transformations. These are given by the following
\bea z &
= & w~e^{f(y^+)/2} \nn \\ x^- & = & y^- - \frac{1}{4} w^2 (\partial_+
f) \nn \\ x^+ & = & y^+~~~~~~~\vx = \vy
\eea
In these coordinates the
metric becomes
\bea ds^2 & = & \frac{1}{w^2} \left[ dw^2 - 2 dy^+ dy^-
+ d\vy^2 + \frac{1}{4} w^2 [(f^\prime)^2 - 2 f^{\prime\prime}] (dy^+)^2
\right]\nn \\ & = & \frac{1}{w^2} \left[ dw^2 - 2 dy^+ dy^- + d\vy^2 +
\frac{1}{4} w^2 (\Phi^\prime)^2 (dy^+)^2 \right]
\label{seven}
\eea
where in the second line we have used (\ref{four}).

The new coordinates provide a new foliation of the space-time.  The
boundary $w = 0$ is naively the same as the original boundary
$z=0$. However, it is well known that AdS/CFT requires an infra-red
cutoff in the bulk which corresponds to a ultraviolet cutoff in the
dual gauge theory. For any such finite cutoff $\epsilon$, the boundary
$w = \epsilon$ is not the same as $z = \epsilon$, and becomes flat in
the $\epsilon \rightarrow 0$ limit. Consequently the dual Yang-Mills
theory lives on a flat space with a $x^+$ dependent coupling.

Notice that in these coordinates, there is only one function $\Phi
(x^+)$ which we are free to choose. In particular, $e^\Phi$ can be
chosen to bounded and vanishing at $x^+=0$ in an {\em analytic}
fashion. The proposed dual will then have a coupling which is bounded
and vanishes at $x^+=0$ in a smooth fashion.

For such solutions, $\partial_+\Phi$ will, however, diverge at
$x^+=0$. This means that the bulk space-time will be as usual
singular. This may be seen by looking at the behavior of geodesics as
in \cite{Das:2006dz,Das:2006pw}. In the new coordinates, these
geodesics are
\bea
w & = & z_0 F(y^+) \nn \\
y^- & = & y^-_0 - \frac{1}{4} z_0^2 \frac{d}{dy^+} (F(y^+)^2 \eea
where $F(y^+)=e^{-f(y^+)/2}$. The affine parameter along such
geodesics is given by \ben \lambda = \int^{y^+} {1 \over
F(y^+)^2}~dy^+ \een It is easy to find functions $\Phi(x^+)$ so
that the singularity at $y^+=0$ is reached in finite affine
parameter. For such colutions $F(y^+)$ must diverge at $y^+ = 0$
The magnitude of the tidal acceleration between two such geodesics
separated along a transverse direction is given by (see equation
(2.10) of \cite{Das:2006pw}) \ben |a| = (F(y^+))^3
F^{\prime\prime} (y^+) \een and would diverge as well. The form of
the metric (\ref{seven}), however, shows that this singularity
weakens as we approach the boundary $w=0$ leaving a flat boundary
metric.

\section{Kasner type solutions}

Similar considerations apply to time dependent solutions of the form
of (\ref{ten}).  We will concentrate on solutions with $p_1=p_2=p_3 =
\frac{1}{3}$. Redefining the time coordinate, this solution may be
written in the form
\bea ds^2 & = & \frac{1}{z^2} \left[ dz^2 + \frac{2t}{3} \left( - dt^2
+ (dx^1)^2 + \cdots (dx^3)^2 \right) \right] \nn \\ e^{\Phi(t)} & = &
|t|^{\sqrt{3}}
\label{nine}
\eea
This solution has a spacelike singularity at $t=0$.

Since the boundary metric on $z=0$ is conformally flat, there
should be a PBH transformation which leads to a foliation with a
flat boundary. This is indeed true. The solution for $t > 0$
becomes \ben ds^2 = \frac{1}{\rho^2}\left[d\rho^2
-\frac{(16T^2-5\rho^2)^2}{256 T^4} dT^2
+\frac{(16T^2-\rho^2)^{\frac{4}{3}}(16T^2+5\rho^2)^{\frac{2}{3}}}{256
T^4} \left( (dx^1)^2 + \cdots (dx^3)^2 \right) \right]
\label{kasnernew}
\een where
the new coordinates $(\rho,T)$ are related to the coordinates
$(z,t)$ in the region $\rho < 4T$ by the transformations \bea z &
= &
\frac{32 \rho T^{\frac{5}{2}}}{\sqrt{6}}\frac{1}{16T^2 - \rho^2} \nn \\
t & = & T~\left( \frac{16T^2 + 5\rho^2}{16T^2-\rho^2}
\right)^{\frac{2}{3}} \label{eleven} \eea The dilaton may be
written down in new coordinates by substituting (\ref{eleven}) in
(\ref{nine}).

 It is clear that in this new
foliation defined by slices of constant
 $\rho$, the boundary $\rho=0$ has
a flat metric. However these coordinate
 system has a coordinate
singularity at $\rho= 4T$, but may be extended
 beyond this point.

The arguments in the previous section then indicate that there is
a dual field theory which lives in a flat space-time, but with a
time dependent coupling which vanishes at $T=0$. Unike the null
solutions,
 the coupling {\em diverges} at early or late times - and
we cannot
 make any careful argument about the behavior of this dual
theory.

As noted in the introduction, these solutions have a curvature
singularity at any finite time, though the singularity goes away at
early and late times. The bulk Ricci scalar is given by
\ben
{\cal R}_5 = -(\frac{9z^2}{4t^3}+20)
\een
In the global geometry the Poincare horizon is a product of a null plane 
times a $S^2$. This singularity appears at {\em one} point on this
null plane. The rest of the Poincare horizon is non-singular.

\section{New class of time dependent solutions}

A necessary condition for a well defined dual theory is that the
coupling should be bounded at all times. This motivates us to search
for new solutions which have space-like singularities of this type.
We will present such solutions in this section.

These solutions are special cases of a class of time dependent
solutions whose boundary metrics are FRW universes. The Einstein frame
metric\footnote{These
solutions can be derived from a generic ansatz with diagonal metric, 
and imposing that the dilaton is a (spatially homogeneous) function 
$\Phi(t)$ of time $t$ alone.} is given by
\ben ds^2=\frac{1}{z^2} \left[ dz^2 +
A(t)[-dt^2+{dr^2\over
1-k\,r^2}+r^2\,(d\theta^2+\sin{\theta}^2\,d\phi^2)] \right]
\label{twelve}
\een
with $k=0,\pm1$, and
\ben
\Phi(t)=\pm \sqrt{3} \int \, {dt \over A(t)}
\een
where
\ben
A(t)=C_1\,\sin({2\,\sqrt{k}\,t})+C_2\,\cos({2\,\sqrt{k}\,t})\ .
\een

The solutions with $k = -1$ are particularly interesting. If we
choose $A(t) = |\sinh (2t)|$, the dilaton becomes \ben e^{\Phi (t)}
= g_s~ |\tanh~t|^{{\sqrt{3}}} \label{twelvea} \een so that the
coupling is bounded and vanishes at $t=0$. There is a spacelike
singularity at $t=0$. In the following we will restrict our attention
to the ``big crunch'' part of the space-time, i.e. for $t < 0$.
In this case we use $A(t) = |\sinh(2t)|$.

The boundary metric is in fact conformal to
parts of Minkowski space. This is seen by defining new coordinates
(for $t < 0$)
\ben
r = \frac{R}{\sqrt{\eta^2 - R^2}}~~~~~~e^{-t} = \sqrt{\eta^2 -
R^2}
\label{thirteena}
\een
The solution now becomes
\bea ds^2 & = & \frac{1}{z^2}
\left[ dz^2 + \mid 1 - \frac{1}{(\eta^2 -
R^2)^2}\mid[ -d\eta^2 + dR^2 + R^2 d\Omega_2^2] \right] \nn \\
e^{\Phi} & = & \mid \frac{\eta^2 - R^2 -1}{\eta^2 - R^2 +1}
\mid^{\sqrt{3}} \label{thirteen}
\eea
The $t >0$ part of the solution also becomes this metric
after a coordinate transformation obtained by reversing the sign of
$t$ in (\ref{thirteena}).
In these coordinates it is clear that as $t \rightarrow -\infty$, i.e.
$\eta^2 - R^2 \rightarrow \infty$, the space-time is $AdS$ and
$e^\Phi$ asymptotes to a constant.

The coordinate
transformation (\ref{thirteena}) is valid in the region $\eta^2 -
R^2 > 0$, and $\eta^2 - R^2 = 1$ are the two spacelike
singularities. Even though we started with the form of the metric
(\ref{twelve}) we could extend the solution beyond part this part
of Minkowski space in the standard manner. In this extended
solution, there are timelike singularities at $R^2 - \eta^2 = 1$.
As is evident, the dilaton shows a singular behavior at the
location of these singularities, even though the value of $e^\Phi$
goes to zero.
In the following we will be interested in the solution in the regions
$(\eta^2 - R^2) > 1$, i.e. the space-time described by the metric
(\ref{twelve}).

Like the Kasner solutions, these solutions generically have curvature
singularities at $z= \infty$.  In the big crunch region, 
The bulk Ricci scalar is given by
\ben {\cal R}_5 = 
-\left( 
20 - 3\frac{z^2}{(\sinh(2t))^3} \right) ,
\een 
where $t$ is as in
(\ref{twelve}) with $A(t) = - \sinh (2t)$ in this $t < 0$ region
of the spacetime. The global nature of this singularity is similar to
the Kasner type solution.  In particular, at early times $t
\rightarrow -\infty$ 
there is no such singularity.

We can, therefore, view these backgrounds in the same way as the null
backgrounds. For $t < 0$ the space-time is pure $AdS_5 \times
S^5$ in the infinite past.
As time evolves one generates a {\em space-like singularity} at
$t=0$ which extend to the boundary defined at $z
=0$. However, since the boundary metric is conformal to flat space, we
can choose a different foliation by performing a PBH transformation
and choose a boundary which is completely flat. (In this case, we have
not been able to find the exact PBH transformations, but - as detailed
in the Appendix - the PBH transformation may be found in an expansion
around the boundary). The gauge theory defined on this latter boundary
is on flat space with a time dependent coupling constant which
vanishes at the location of the bulk singularity. The source in the
gauge theory evolves the initial vaccum state. On the supergravity
side, a (timelike) singularity develops at $z= \infty$. While we not
have a clear idea of the meaning of this singularity in the gauge
theory it is reasonable to presume - in view of the usual AdS/CFT
duality - that this should manifest itself in the infrared
behavior. Finally, as the time evolves, the gauge coupling goes to
zero - this manifests itself as a spacelike singularity in the bulk in
a region where supergravity itself breaks down.

The analysis of this dual gauge theory appears to be more complicated
than the dual gauge theory for null backgrounds. One issue is related
to the fact that the gauge theory lagrangian has an overall factor of
$e^{-\Phi}$.
When $\Phi$ depends only on a null direction, it
was shown in \cite{Das:2006pw} that a choice of light cone gauge, together
with a field redefinition converts the kinetic terms in the action
into standard form for constant couplings. All factors of couplings
then appear in the nonlinear terms as positive powers of $e^{\Phi
  (x^+)}$, which vanish at the location of the bulk singularity. This
allowed us to arrive at some clean conclusions about the behavior of
the gauge theory. In \cite{Chu:2007um} a different gauge choice was used -
this again made analysis of the gauge theory easier. For time
dependent backgrounds, we have not been able to find a gauge choice
and a field redefition which leads to a similar
simplification. Nevertheless we expect that the theory is amenable to
perturbative analysis near $t = 0$ where the gauge coupling becomes
weak.

\section{ The Holographic Stress Tensor}

In this section we use the standard
techniques of covariant counterterms
\cite{skenderis,balasubramanian,nojiri:2000,Emparan:1999pm,Awad:1999xx}
to calculate the holographic stress tensor. The gravity-dilaton action in
five dimensional space ${\cal M}$, with boundary $\partial {\cal M}$
is given by,
\begin{eqnarray}
I_{\rm bulk}+I_{\rm surf}= {1 \over 16 \pi G_5}\int_{\cal M}
d^{5}x \sqrt{-g}\left( R^{(5)}+{12}-{1\over 2}(
\nabla\,\Phi)^2\right)-{1 \over 8 \pi G_5 } \int_{\partial {\cal
M}} d^{4}x \sqrt{-h} \,\Theta.
\label{mainaction}
\end{eqnarray}
Where the second term is the Gibbons-Hawking boundary term,
$h_{\mu\nu}$ is the induced metric on the boundary and $\Theta$ is
the trace of the extrinsic curvature\footnote{$\Theta_{a
b}=-{1\over 2}(\nabla_a\,n_{b}+\nabla_b\,n_a)$, where $n^a$ in the
unit normal vector to the surface z=constant and pointing to the
boundary $\partial{\cal M}$} of the boundary $\partial{\cal M}$.

The above action is divergent. Therefore, one might use one of the
known techniques to regularize such action. Here we choose to work
with the covariant counterterm method since we are interested in
calculating the boundary energy momentum and its trace. To have a
finite action one can add the following counterterms
\begin{eqnarray} I_{\rm ct}= - {1 \over 8 \pi G_5} \int_{\partial {\cal
M}}d^{4}x\sqrt{-h}\Biggl[ {3}+{{\cal R} \over 4}-{1 \over
8}(\nabla\,\Phi)^2 - \log(\rho_0)~a_{(4)}\Bigg]\, \label{theterms}
\end{eqnarray}
where $\rho_0$ is a cutoff on the radial coordinate $\rho$ which has
to be taken to zero at the end of the calculation.
${\cal R}$ is the Ricci
scalar for $h$. The term proportional to $\log (\rho_0)$ is required
to cancel a logarithmic divergence in the action
(\ref{mainaction}). However this term does not contribute to the
renormalized energy momentum tensor.

Now the total action is given by $I{=}I_{\rm
bulk}{+}I_{\rm surf}{+}I_{\rm ct}$. Using this action one can
construct a divergence free stress energy tensor
\cite{balasubramanian}:
\begin{eqnarray}  T^{\mu\nu}&=& {2 \over \sqrt{-h}} {\delta I
\over \delta h_{\mu\nu}}\nonumber\\
&=&{1 \over 8 \pi G_5} \Biggl[
\Theta^{\mu\nu}-\Theta\,h^{\mu\nu}-3\,{h^{\mu\nu}}+{1 \over
2}G^{\mu\nu}-{1\over
4}\nabla^{\mu}\,\Phi\nabla^{\nu}\,\Phi+{1\over
8}h^{\mu\nu}(\nabla\,\Phi)^2\Bigg]\, \label{stressone}
\end{eqnarray}
Here $G_{\mu\nu}$ and $\nabla$ are the Einstein tensor and
covariant derivative with respect to h. In the regime where the
supergravity approximation is valid, the vev of the CFT's
energy momentum tensor ${< T^{\mu\nu}>}$ is related the above
stress tensor by the following equation
\begin{equation}
 \sqrt{-\tilde{g}}\,\tilde{g}_{\mu\nu}{<
T^{\nu\sigma}>}=\lim_{z\to
 0}\sqrt{-h}\,h_{\mu\nu}T^{\nu\sigma}\ . \label{newstress}
\end{equation}
where we have used the notation of equation (\ref{fgform}).

The energy momentum tensors calculated in the holographic RG approach
correspond to operators in the dual field theory which are regularized
using the specific boundary metric used to perform the bulk
calculation.

\subsection{Conformally Flat Boundary}

Let us first consider bulk metrics of the form (\ref{one}). This means
we use a cutoff defined in terms of the radial coordinate $z$.
Using the above expression for the stress tensor,
one can easily show that
for a any solution with conformally flat boundary
(i.e. of the form of equation (\ref{two}), the stress
tensor vanishes. Let us see how this result is obtained. First,
the extrinsic curvature for a solution with a conformally flat
boundary is
\ben \Theta_{\mu\nu} = -\,h_{\mu\nu}.
\een
The extrinsic
curvature terms in the expression then cancel with the term
proportional with the induced metric. Using (\ref{three})
and its contraction, one
can see directly that the last three terms exactly cancel leading
to the vanishing of the stress tensor. As a result, the trace anomaly
vanishes. Comparing this result with the known results in the
literature one finds the following. Our result does not match with
the field theory calculation of trace anomaly in
\cite{Liu:1998bu}. The reason is that in this calculation only
terms up to quadratic order in the dilaton were included and all
higher orders have been ignored. But this result agrees with the
holographic anomaly expression calculated in \cite{nojiri:2000}
since their expression contains these terms which are crucial to
have a vanishing anomaly.

\subsection{Flat boundary}

We will now consider the energy momentum tensor which is defined by a
choice of foliation which leads to a {\em flat} boundary metric. This
is of course a different regularization and would lead to a different
answer which would give us the energy momentum tensor of the gauge
theory defined on flat space in an appropriate regime.

\subsubsection{Null solutions}
It is easy to check by a direct calculation
that for the solutions with null singularities,
the energy momentum tensor continues to vanish.

\subsubsection{Kasner-type solutions}
Now consider the Kasner type solution in new coordinates
(\ref{kasnernew}).
Using the above expression for the
holographic stress tensor, one gets the following \ben
T_{\mu}^{\nu}={ \rho^4 \over 1024\,\pi\,G_5\, T^4}\, \mbox {diag}
\,(\,9,\, 13,\, 13,\, 13)+\,\,O(\rho^6). \een The energy momentum
tensor of the CFT as in (\ref{newstress}) is given by \ben
<T_{\mu}^{\nu}>={N^2 \over 512 \,\pi^2\, T^4}\,\mbox {diag}
\,(\,9,\, 13,\, 13,\, 13), \een which has the following
non-vanishing trace: \ben <T_{\mu}^{\mu}>={3\,N^2 \over 32\,
\pi^2}{1 \over T^4}, \een here we have used \ben G_5={\pi \over
2\,N^2}. \label{defg5} \een
The trace computed here agrees with
the holographic trace anomaly found in \cite{nojiri:2000}.
The nonzero energy momentum tensor can be possibly interpreted
as particle production.

\subsubsection{FRW solutions}

To calculate the energy momentum tensor for the new FRW solution
with $k=-1$ it is convenient to work with the following
coordinate system. These coordinates allow the conformal factor
to depend only on one coordinate, while keeping the boundary
metric Minkowski. The coordinate transformations are as follows:
\ben
\tau^2=\eta^2-R^2,\hspace{0.5in}r^2={\eta+R \over \eta-R}
\een This puts the metric in the form \ben ds^2={dz^2 \over
z^2}+{1\over z^2}\,(1-{1\over \tau^4})\,[-d\tau^2+{\tau^2 \over
r^2}\,dr^2+{\tau^2 \over 4}\,(r-{ 1\over r})^2\, d {\Omega_2}^{2}]
\een The dilaton in this coordinates is given by \ben \Phi(\tau)
=\sqrt{3}\, \ln \left[{\tau^2-1 \over \tau^2+1} \right]
\een
One
can use this coordinates to do a PBH transformations as explained
in the appendix and obtain another form of this solution with
Minkowski boundary. In this form the stress energy tensor is given
by \ben T_{\mu}^{\nu}={ \rho^4 \over 4\,\pi\,G_5\, (\bT^4-1)^4}\,
\mbox {diag}
\,\left(12-3\,\bT^4,\,4+9\,\bT^4,\,4+9\,\bT^4,\,4+9\,
\bT^4\right)+\,\,O(\rho^6).
\een
where the coordinate $\bT$ is defined in the Appendix. Using
(\ref{newstress}) and (\ref{defg5}), the energy momentum tensor of
the CFT is given by \ben <T_{\mu}^{\nu}>={ N^2 \over
2\,\pi^2\,(\bT^4-1)^4}\, \mbox {diag} \,\left(
12-3\,\bT^4,4+9\,\bT^4,4+9\,\bT^4,4+9\,\bT^4 \right),
\label{stensorflat}
\een which
has the following non-vanishing trace: \ben
<T_{\mu}^{\mu}>={12\,N^2 \over \pi^2}{(\bT^4+1) \over (\bT^4-1)^4}
\een Again this trace agrees with the calculation in
\cite{nojiri:2000}.
Note that the energy momentum tensor vanishes at early times. This
reinforces our claim that at early times we have started with The
vacuum state of the dual gauge theory, with a source which vanishes 
at $\bT \rightarrow -\infty$. At later times, the source produces 
a nonzero energy momentum tensor as well as a nonzero expectation 
value of the operator dual to the dilaton. In other words, the Heisenberg
picture state is the vacuum of the CFT.
It is tempting to interpret the nonzero stress tensor as an effect of
particle production.
Once again the stress tensor diverges at the singularity
${\bar{T}}=1$. However this is precisely the place where the
holographic calculation cannot be trusted.

The real question is whether the gauge theory is well behaved in this
region. For null backgrounds, this appears to be so
\cite{dasun,Chu:2007um,Lin:2006sx}
For time dependent backgrounds, this is not
clear at the moment, particularly because of bulk singularities at $z
= \infty$ which signify that there are infrared problems in the gauge
theory. These issues are under investigation.

\section{Acknowledgements}

We would like to thank Costas Bachas, Ian Ellwood, Ben Craps,
David Gross,
A. Harindranath, David Kutasov, Gautam Mandal, Shiraz Minwalla,
Tristan McLoughlin and Alfred Shapere for discussions at various
stages of this work and Steve Shenker for a correspondence.
S.R.D. would like to thank Tata Institute of
Fundamental Research, Indian Association for the Cultivation of
Science and Bensque Center for Science for hospitality. Some of the
results were presented by S.R.D. in a talk at Benasque String Workshop
in July 2007.
K.N. would like to thank TIFR for hospitality during
various stages of this work, as well as the organizers of Strings 07,
Madrid, and the String Cosmology workshops at ICTP, Trieste and KITPC,
Beijing, where some of this work was done. S.T. thanks the Swarnajayanti 
Fellowship, DST. Govt. of India, for support. 
The research reported
here was supported in part by the United States National Science
Foundation Grant Numbers PHY-0244811 and PHY-0555444 and Department of
Energy contract No.  DE-FG02-00ER45832, as well as the Project of Knowledge
 Innovation Program (PKIP) of the
Chinese Academy of Sciences. It was also supported by the DAE, 
Govt. of India, and especially the people of India, whom we  thank.

\appendix
\section{PBH transformations}

In the coordinates displayed in (23), the spacelike singularity
extends to the boundary. However in the form (\ref{thirteen}) the
boundary metric is conformal to flat space. This suggests that
there should be PBH transformations which leads to a flat boundary
metric. In the case of FRW solutions however, we have not yet been
able to find the exact PBH transformations. We will show below how
to find these transformations systematically in the neighborhood
of the boundary and obtain them to the order which is required for
our analysis of the energy momentum tensor in the next section.

Let us show how can we obtain such coordinate transformation for a
solution with a conformally flat boundary on the following form
\ben ds^2={1\over z^2}\,[dz^2+
f(t)\,\eta_{\mu\nu}\,dx^{\mu}dx^{\nu}] \een We chose the conformal
factor to depend only on one coordinate since this will be
sufficient to deal with the cases under consideration in this
work. One can generalize such a procedure to cases with general
conformal factor $f(x^{\mu})$ and boundaries other Minkowski.
Define the following coordinate transformations \bea
t(\rho,T)&=&\sum_{n=0,2,..}\, a_i(T)\,\rho^{i}\nonumber\\
z(\rho,T)&=&\rho\,\sum_{n=0,2,..}\, s_i(T)\,\rho^{i} \eea One can
choose $a_0(T)=T$, then expanding all metric components in $\rho$,
they have the following form in new coordinates \bea
g_{\rho\rho}&=& {1 \over \rho^{2}}+{4\over
s_o^2}[\,s_0\,s_2-a_2^2\,f\,]+{ 4 \over
s_0^3}\,[\,2\,s_4\,s_0^2-a_2^3\,\dot{f}\,s_0+2\,a_2^2\,f\,s_2-4\,a_2\,f\,a_4\,s_0\,]\,\rho^2
+O(\rho^4)\nonumber\\
g_{\rho T}&=&{1 \over s_0^2}[\,\dot{s}_0\,s_0-2\,a_2\,f\,] \, {1
\over \rho}+ {1 \over
s_0^3}\,[\dot{s}_2\,s_0^2+s_2\,\dot{s}_0\,s_0-2\,a_2\,f\,\dot{a}_2\,s_0
-2\,a_2^2\,\dot{f}\,s_0\nonumber\\&&+4\,f\,a_2\,s_2-4\,f\,a_4\,s_0\,]\,\rho+O(\rho^3)\nonumber\\
g_{TT}&=&{-f \over s_0^2}\,{1 \over \rho^{2}}-{1\over
s_0^3}\,[a_2\,s_0\,\dot{f}
-2\,f\,s_2+2\,\dot{a}_2\,f\,s_0-\dot{s}_0^2s_0\,]-{1 \over
2\,s_0^4}[2\,s_0^2\,\dot{f}\,a_4+s_0^2\,\ddot{f}\,a_2^2
-4\,\dot{f}\,a_2\,s_0\,s_2\nonumber\\&&+4\,\dot{f}\,a_2s_0^2\,\dot{a}_2-4\,f\,s_4\,s_0+6\,f\,s_2^2-8\,f\,
\dot{a}_2\,s_2\,s_0
+4\,f\,s_0^2\,\dot{a}_4+2\,f\,s_0^2\,\dot{a}_2^2\nonumber\\
&&+4\,\dot{s}_0^2\,s_2\,s_0-4\,\dot{s}_0\,\dot{s}_2s_0^2]\,\rho^2+O(\rho^4)\nonumber\\
g_{ii}&=&{f \over s_0^2}\, {1 \over \rho^{2}}-{1\over
s_0^3}\,[2\,s_2\,f-a_2\,s_0\,\dot{f}]+ {1 \over
2\,s_0^4}\,[2\,s_0^2\,\dot{f}\,a_4+s_0^2\,\ddot{f}\,a_2^2-4\,\dot{f}\,a_2\,s_0\,s_2\nonumber\\
&&-4\,f\,s_4\,s_0+6\,f\,s_2^2]\,\rho^2+O(\rho^4) \eea To keep the
PG form of the metric and to get a Minkowski boundary, one imposes
the following conditions \ben g_{\rho\rho}={ 1 \over
\rho^{2}},\hspace{.75in} g_{\rho T}=0,\hspace{.75in}
g_{\mu\nu}=\eta_{\mu\nu}\,{ 1 \over \rho^{2}}\,+O(1),\een this
guarantee the existence of such a coordinate system, at least,
close to the new boundary. These conditions lead to \bea
s_0(T)&=&f(T)^{1 \over 2},\hspace{.25in}\,s_2(T)={\dot{f}(T)^2
\over 16\,f(T)^{3 \over 2}}\hspace{.25in}\,s_4(T)={\dot{f}(T)^4
\over 256\,f(T)^{7 \over 2}},\,..\nonumber\\
a_2(T)&=&{\dot{f}(T) \over 4 f(T)},\hspace{.25in}\,a_4(T)=0,\,..
\eea Applying the above procedure to Kasner type solutions in (20)
with $f(t)={2 \over 3}\,t$, one can obtain the following
coordinate transformations \ben z(\rho,T)={\sqrt{6\,T}\, \rho
\over 3}\,[1+ {\rho^2 \over 16 T^2}+{\rho^4 \over 256
T^4}]+O(\rho^7), \hspace{0.3in} t(\rho,T)=T+{\rho^2\over
4\,T}+O(\rho^6) \een

The metric in these coordinates has a Minkowski boundary and has
the following form \ben ds^2=[{1 \over \rho^2}+O(\rho^4)]d\rho^2
-[{ 1 \over \rho^{2}}\,-{ 5 \over 8\,T^2}+{25\over
256\,T^4}\,\rho^2+O(\rho^4)]\,dT^2+[{ 1 \over \rho^{2}}\,+{ 1
\over 8\,T^2}-{7\over 256\,T^4}\,\rho^2+O(\rho^4)]\,d\bar{x}^2\een
which agrees with the exact coordinate transformation in (22) and
the metric (21) upon expanding it in powers of $\rho$.

Before applying this procedure to calculate the FRW solution with
$k=-1$ let us use the coordinate system given in (38). \ben
\tau^2=\eta^2-R^2,\hspace{0.5in}r^2={\eta+R \over \eta-R}\een This
puts the metric in the form \ben ds^2={dz^2 \over z^2}+{1\over
z^2}\,(1-{1\over \tau^4})\,[-d\tau^2+{t^2 \over r^2}\,dr^2+{\tau^2
\over 4}\,(r-{ 1\over r})^2\, d {\Omega_2}^{2}]
 \een
Following the above procedure one can obtain the PG form of this
solution with Minkowski boundary. The coordinate transformations
and the metric are
 \bea
z(\rho,\bT) & = & {\sqrt{\bT^4-1} \rho\over
\bT^2}\,[1+ {\rho^2 \over \bT^2\,(\bT^4-1)^2}+{\rho^4 \over
\bT^4\,(\bT^4-1)^4}]+O(\rho^7), \nn \\
\tau(\rho,\bT) & = & \bT+{\rho^2\over (\bT^4-1)\,\bT}+O(\rho^6) 
\eea

\bea ds^2&=&[{ 1 \over \rho^{2}}\,+O(\rho^4)]\,d{\rho}^2-[{ 1
\over \rho^{2}}\,-{10\,\bT^2 \over (\bT^4-1)^2}+{25\,\bT^4 \over
(\bT^4-1)^4}\,\rho^2+O(\rho^4)]\,d\bT^2\nonumber\\
&+&[{ 1 \over \rho^{2}}\,+{2\,\bT^2\over
(\bT^4-1)^2}+{(\bT^4-8)\over
(\bT^4-1)^4}\,\rho^2+O(\rho^4)]\,\left[{\bT^2\over
r^2}\,dr^2+{\,\bT^2\over 4}\,(r-{1 \over
r})^2\,d\Omega_2^2\right]\eea

\end{document}